\documentclass{emulateapj}
\usepackage{apjfonts}
\submitted{Accepted to ApJ}

\usepackage{amssymb}
\usepackage{amsmath}
\usepackage{graphicx}

\newcommand{\mbfB}{\mathbf{B}}

\newcommand{\om}{\omega}
\newcommand{\omce}{\omega_{ce}}
\newcommand{\omci}{\omega_{ci}}
\newcommand{\ompe}{\omega_{pe}}
\newcommand{\ompi}{\omega_{pi}}
\newcommand{\ncr}{n_{cr}}
\newcommand{\rcr}{r_{cr}}
\newcommand{\ri}{r_i}
\newcommand{\kb}{k_{Bell}}
\newcommand{\kw}{k_{wice}}

\begin{document}

\title{Environments for Magnetic Field Amplification by Cosmic Rays}

\author{Ellen G. Zweibel\altaffilmark{1,2}}
\author{John E. Everett\altaffilmark{1,2}}
\altaffiltext{1}{U Wisconsin-Madison, 475 N Charter St, Madison,
                 WI 53706, U.S.A.}
\altaffiltext{2}{Astronomy \& Physics Depts.
\& Center for Magnetic Self-Organization in Laboratory \& Astrophysical
Plasmas}

\shorttitle{Zweibel \& Everett}
\righthead{Weak Magnetic Fields}

\begin{abstract}
We consider a recently discovered class of instabilities, driven by
cosmic ray streaming, in a variety of environments.  We show that
although these instabilities have been discussed primarily in the
context of supernova driven interstellar shocks, they can also operate
in the intergalactic medium and in galaxies with weak magnetic fields,
where, as a strong source of helical magnetic fluctuations, they could
contribute to the overall evolution of the magnetic field. Within the Milky
Way, these instabilities are strongest in warm ionized gas, and appear
to be weak in hot, low density gas unless the injection efficiency of cosmic
rays is very high.
\end{abstract}

\section{Introduction}\label{s:introduction}

Recently a powerful instability which couples a high flux of cosmic
rays to their host medium has been discovered (Bell 2004, Blasi \&
Amato 2008); we refer to this as the ``Bell Instability.''  The
instability amplifies low frequency, right circularly polarized
electromagnetic fluctuations with wavenumber parallel to the ambient
magnetic field. In contrast to the classical cyclotron resonant
streaming instability (Wentzel 1968, Kulsrud \& Pearce 1969), in which
cosmic rays with Lorentz factor $\gamma$ amplify Alfv\'en waves with
wavelength of order the cosmic ray gyroradius $r_{cr}\sim \gamma
c/\omci$ (where $\omci$ is the non-relativistic ion-cyclotron
frequency), the characteristic wavelength of the Bell instability is
much less than $r_{cr}$.  The Bell instability is thought to be an
important ingredient of diffusive shock acceleration in supernova
remnants. It might be responsible for amplifying the magnetic field by
up to $\sim 2$ orders of magnitude above its interstellar value and
increasing the maximum energy to which cosmic rays can be accelerated,
possibly up to the ``knee" at $\sim 10^{15}$\,eV (Drury 2005, Reville
et al. 2008a). This is an important result, because it has been known
since the work of Lagage \& Cesarsky (1983) that standard models of
diffusive shock acceleration in the interstellar medium fail to reach
the energy of the knee.

Although the Bell instability has been applied primarily to supernova
remnants, there are many other environments in which the cosmic ray
flux may be large enough to excite it: galactic wind termination
shocks, intergalactic shocks, and shocks in disks and jets. Cosmic
rays streaming away from local sources or from their host galaxies may
also constitute a sufficient flux. If the Bell instability exists in
any of these environments it can amplify the magnetic field and
transfer cosmic ray energy and momentum to the background plasma as
well as increasing the efficiency of cosmic ray acceleration. In view
of our current uncertainty as to how galactic and intergalactic
magnetic fields originated and are maintained (Widrow 2002, Kulsrud \&
Zweibel 2008), any mechanism for amplifying them is of interest.
Because fluctuations generated by the Bell instability have a definite
helicity relative to the background magnetic field, they could be
significant in amplifying the field at large scales (Pouquet et al.
1976).  This paper assesses the conditions under which the Bell
instability, or more generally any rapidly growing, nonresonant
electromagnetic streaming instability, can exist.

In order to excite the Bell instability, the cosmic ray particle flux
$\ncr v_D$, thermal ion density $n_i$, and Alfven speed $v_A$ must
satisfy the inequality (as we show in \S\ref{s:stdCase})
\begin{equation}\label{bellineq}
\ncr v_D > n_i\frac{v_A^2}{\langle\gamma\rangle c},
\end{equation}
where $\langle\gamma\rangle$ is of order the mean cosmic ray Lorentz
factor. This can be written in terms of the cosmic ray and magnetic
energy densities $U_{cr}$, $U_B$,
\begin{equation}\label{ucrub}
\frac{U_{cr}}{U_B}>\frac{c}{v_D}.
\end{equation}
Equations (\ref{bellineq}) and (\ref{ucrub}) express the requirement that
the characteristic wavenumber of the Bell instability be much greater than
the reciprocal of the mean cosmic ray gyroradius $c\langle\gamma\rangle/\omci$.

The Bell instability was originally derived for a cold plasma, which
is valid when the thermal ion and electron gyroradii $r_i$, $r_e$, are
much less than the characteristic wavelength $\kb$ of the instability;
$\kb r_{i,e}\rightarrow 0$.  Reville et al. (2008b) considered $kr_i$
small but nonzero. Their condition that the instability is
significantly modified by thermal effects can be written in terms of
the cosmic ray flux and ion thermal velocity $v_i\equiv
\sqrt{2k_BT/m_i}$ are (as we derive in \S\ref{s:wice})
\begin{equation}\label{revilleineq}
\ncr v_D > n_i\frac{v_A^3}{v_i^2}.
\end{equation}
When the inequality (\ref{revilleineq}) is satisfied, the wavenumber
of maximum instability, $\kw$ (warm ions, cold electrons) decreases
relative to $\kb$. And, while the maximum growth rate of the Bell instability,
and the growth rate of the
resonant streaming instability are independent of $B$ as long
as $v_D$ much exceeds the Alfven speed $v_A$, the growth rate of the
thermally modified instability increases linearly with $B$.

\begin{figure*}[ht]
\begin{center}
\includegraphics[height=100mm]{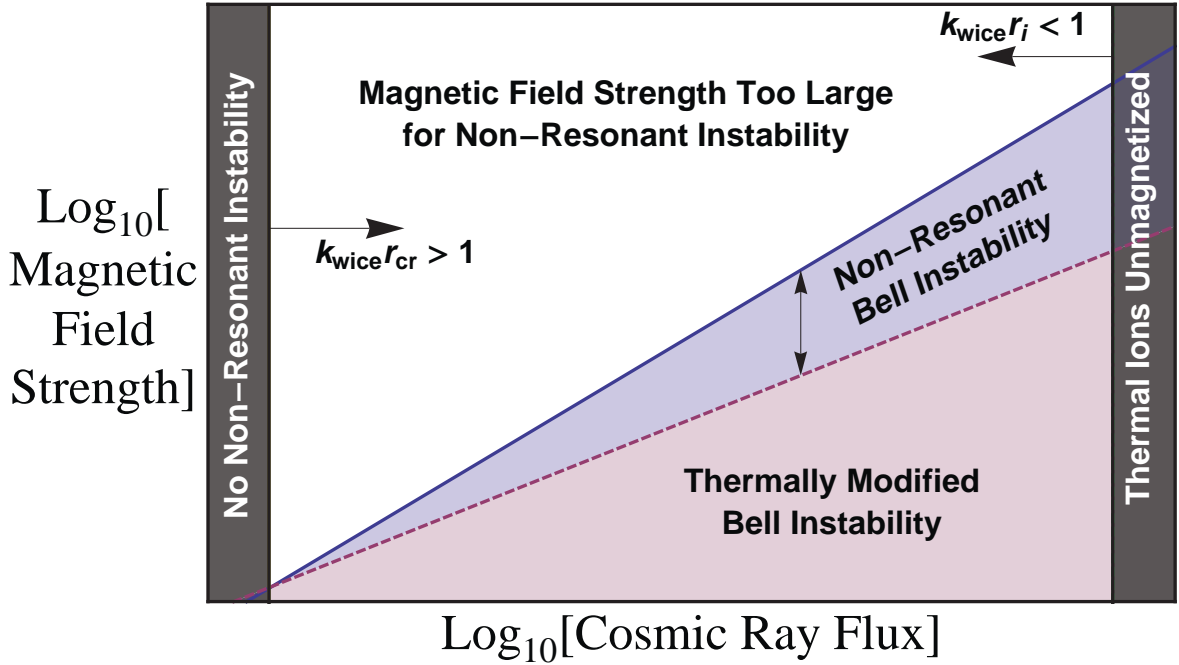}
\caption{A schematic of the instability regimes in the
cosmic-ray flux vs. magnetic-field strength plane.  At high
magnetic-field strengths, in the unshaded region in this figure, the
magnetic field is too strong to allow the non-resonant Bell
Instability to grow.  At lower magnetic-field strengths, a cosmic-ray
flux dependent threshold (shown here with the solid line) is reached
where the cosmic rays can launch the Bell instability; this limiting
magnetic field scales as $(n_{cr} v_D)^{1/2}$.  At lower
magnetic-field strengths (below the dashed line in this figure),
thermal-pressure effects become important, and modify the maximum
growth-rate of the Bell Instability; this threshold scales as $(n_{cr}
v_D)^{1/3}$.  All of these regions are bounded on the left and the
right by limits on the cosmic-ray flux: to the left ($\kw\rcr > 1$) is
the limit that the cosmic-ray flux must be high enough to excite the
non-resonant Bell Instability.  To the right ($\kw\ri < 1$) is the
limit beyond which the thermal ions are no longer magnetized; i.e.,
the gyroradius of the thermal ions is larger than the wavelength of
the instability.  The quantitative labels on this plane change with $\langle\gamma\rangle$,
$T$, and $n_i$; quantitative results are given for particular physical
cases, with $\langle\gamma\rangle = 1$, in Figures~\ref{ismregimes.eps} and \ref{icmregimes.eps}.
\label{planesketch.eps}}
\end{center}
\end{figure*}

In this paper we extend Reville et al.'s analysis to cases where
$kr_i$ is not small, include cyclotron damping, and investigate the
properties of the instability by solving the plasma dispersion
relation. The results are given schematically in
Figure~\ref{planesketch.eps} and precisely for two representative
environments, for $\langle\gamma\rangle = 1$, in
\S\ref{ss:regimes}. Equations (\ref{bellineq}) and (\ref{revilleineq})
define curves on the ($\ncr v_D,B$) plane.  We show that these two
curves, together with the requirements $\kw r_{cr}> 1$, $\kw r_i < 1$,
and $\kb r_i < 1$, divide the $(\ncr v_D, B)$ plane into the domains
delineated in the figure. First, we consider the case of $\kw\rcr >
1$: if the cosmic ray flux is too low, resonant streaming
instabilities can be excited, but nonresonant instabilities of the
Bell or thermally modified Bell type are precluded. The threshold flux
depends only on $v_i$ and the ion density $n_i$, and is independent of
$B$. Above this threshold, but below a second threshold, defined by
$\kw r_i < 1$, the plane is divided into three regions. If $B$ is
large enough that eqn. (\ref{bellineq}) is violated, nonresonant
instability is again precluded.  Intermediate values of $B$, such that
eqn. (\ref{revilleineq}) is violated but eqn. (\ref{bellineq}) is not,
define the range of the standard Bell instability. If $B$ is small
enough that eqn. (\ref{revilleineq}) holds, there is nonresonant
instability of the thermally modified type. The condition $\kw r_i \le
1$ gives an upper limit to the flux at which nonresonant instability
can be excited. Above this flux, the ions are unmagnetized and any
instability which exists is driven solely by the electrons.  This
limit too, which is discussed further in \S 2 and in the Appendix, is
independent of $B$ and depends only on $n_i$ and $v_i$.  Evaluating
these limits numerically, we find that there is a broad range of
cosmic ray flux and magnetic fieldstrength for which the instability
operates efficiently, should mediate the transfer of cosmic ray
momentum and energy to the background, and should be a powerful source
of circularly polarized magnetic field fluctuations.

In \S\ref{s:setup} we write down the general dispersion relation and
solve it in various limits. In \S\ref{s:applications} we give
quantitative versions of Figure~\ref{planesketch.eps} and
apply the results to several different astrophysical
environments. Section \ref{s:summary} is a summary and
discussion. Gaussian cgs units are used throughout.

This paper is restricted to linear theory. The quasilinear and
nonlinear evolution of the instability has been studied by a number of
authors using both magnetohydrodynamic (Bell 2004, 2005, Pelletier et
al. 2006, Zirakashvili et al. 2008) and kinetic models (Niemiec et
al. 2008, Riquelme \& Spitkovsky 2009, Ohira et al. 2009, Luo \&
Melrose 2009, Vladimirov et al.  2009). A host of saturation
mechanisms have been investigated, including increase of the
fluctuation scale to $r_{cr}^{-1}$, acceleration of the thermal
plasma, coupling to stable or damped modes, and relaxation of the
exciting beam. Although understanding saturation is essential for
predicting the outcomes of the instability, delineating the regimes in
parameter space where the instability exists is the first step.
\section{Derivation of the Instability}\label{s:setup}

We analyze the situation considered by Bell; a singly ionized plasma
with ion number density $n_i$ and temperature $T$. There is a uniform
magnetic field $\mbfB =\hat z B$ and a population of proton cosmic
rays with number density $n_{cr}$ streaming along $\mbfB$ with speed
$v_D$ relative to the thermal ions. We assume from now on that the
mean Lorentz factor $\langle\gamma\rangle$ of the cosmic rays is order
unity.  There are applications, such as a ``layered" shock precursor
in which the most energetic particles have penetrated furthest
upstream, where locally $\langle\gamma\rangle\gg 1$, and our results
can be scaled readily to this situation; that application is important
for determining the maximum energy to which particles can be
accelerated.  However, the instability will be excited by the bulk
cosmic-ray population closer to the shock, which is important for
field amplification.

There are two populations of electrons, one with density $n_i$ which
has no bulk velocity in the frame of the protons, and the other with
density $\ncr$ which drifts with the cosmic rays at speed $v_D$. Thus,
the system is charge neutral and current free.  This is the model used
in Zweibel (2003) and Bell (2004).  Amato \& Blasi (2009) have
considered the Bell instability when all the electrons drift at speed
$(\ncr/n_i)v_D$ and found results similar to those obtained for the
two electron populations assumed here. But although the Bell
instability in its original form is insensitive to the precise form of
the thermal electron distribution function $f_e(v)$, it does turn out
to depend on $f_e$ in a hot plasma.

We briefly consider the constraints on $f_e$ in the Appendix. Based
on an assessment of the electrostatic Langmuir instability, we argue that
for very large drifts and high cosmic ray densities, such as are expected in
young supernova remnants expanding into diffuse interstellar gas, a separate
electron beam with density $\ncr$ and drift velocity $v_D$ is unstable, and would tend to relax.
For more moderate shocks, or in outflows where the drift speed is less than the electron
thermal velocity, such a beam is stable. In general, stability considerations alone do not determine $f_e$.

The fastest growing nonresonant
cosmic rays streaming instabilities are
sensitive to the precise form of $f_e$ only 
in a plasma so hot and/or weakly magnetized that
the thermal ions do not respond. Any instability present is then an
instability of the electrons alone. Its physical significance is
unclear, since it depends on the form of $f_e$.

\subsection{Full Dispersion Relation}
\begin{figure}[h!]
\begin{center}
\includegraphics[height=50mm]{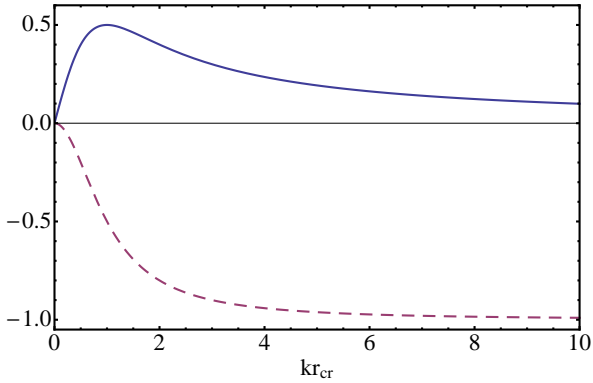}
\caption{\label{zeta.eps} Imaginary part (top panel) and real part
(bottom panel) of the function $\zeta_r$ for right circularly
polarized waves, plotted {\textit{vs}} dimensionless wavenumber
$kc/\omci$, for the normalized cosmic ray distribution function
$\phi(p) = (4/\pi p_0^3)(1+p^2/p_0^2)^{-2}$, where $p_0 = m_pc$.}
\end{center}
\end{figure}

We are interested in right circularly polarized electromagnetic
fluctuations which propagate parallel to $\mbfB$ and depend on $z$ and
$t$ as $\exp{i(kz -\om t)}$; thus instability corresponds to
$Im(\om)\equiv\om_i>0$. The dispersion relation for the fluctuations
can be written as
\begin{eqnarray}\label{fulldr}
\frac{c^2k^2}{\om^2}-\omci\frac{\ncr}{n_i}\zeta_r\frac{\left(\om-kv_D\right)}{\om^2}\frac{c^2}{v_A^2}&=&\frac{\ompi^2}{\om kv_i}
Z\left(\frac{\omci+\om}{kv_i}\right) + \nonumber \\ & & \frac{\ompe^2}{\om kv_e}
Z\left(\frac{\omce+\om}{kv_e}\right),
\end{eqnarray}
where we have dropped the displacement current, $\omega_{pe,i}\equiv (4\pi n_ie^2/m_{e,i})^{1/2}$ are
the thermal electron and ion plasma frequencies, $\omega_{ce,i}\equiv
\mp eB/m_{e,i}c$ are the electron and ion cyclotron frequencies, $ Z$
is the plasma dispersion function (Fried \& Conte 1961)
\begin{equation}\label{Z}
Z(z)\equiv\frac{1}{\sqrt{\pi}}\int_{-\infty}^{\infty}\frac{e^{-s^2}}{s-z}ds,
\end{equation}
and the quantity $\zeta_r$ is defined in eqn. (A10) of (Zweibel 2003)
and is plotted in Figure (\ref{zeta.eps}).
Although the exact behavior of $\zeta_r$ depends on the cosmic ray
distribution function, it has the general property that when $k\rcr\gg
1$, $\zeta_r \rightarrow -1$, with a small imaginary part of order
$(k\rcr)^{-1}$. The physics underlying this behavior is that cosmic
rays barely respond to disturbances with wavelengths much less than
their gyroradius, but the electrons do respond, resulting in a large
perturbed current. A fraction $(k\rcr)^{-1}$ of cosmic rays have large
enough pitch angles that they can resonate, resulting in a small
imaginary part. As to the background plasma terms, the first term on
the right hand side of eqn. (\ref{fulldr}) represents the response of
the thermal ions, and the second term represents the cold electrons.

\begin{figure}[h!]
\begin{center}
\includegraphics[height=50mm]{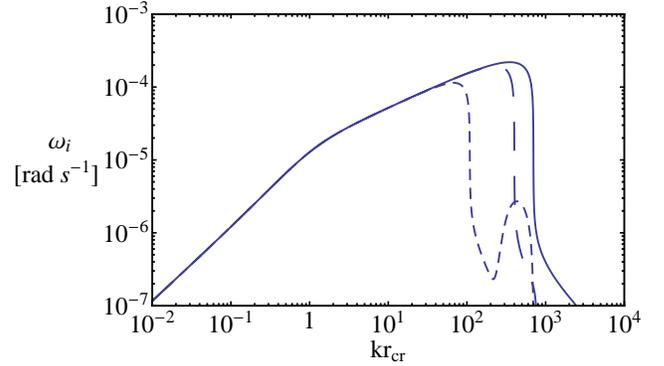}
\caption{\label{growthrates.eps}Growth rates $\om_i$ {\textit{vs}}
wavenumber $k$ in units of $r_{cr}^{-1} \equiv\omci/c$ for three
different temperatures in a medium with $B=3\,\mu G$, $n_i=1$\,cm$^{-3}$,
$\ncr v_D=10^4$\,cm$^{-2}$\,s$^{-1}$. Solid line: $T=10^4$\,K, long dashed
line: $T=10^6$\,K, short dashed line: $T=10^7$\,K.}
\end{center}
\end{figure}

We have solved eqn. (\ref{fulldr}) for a variety of ambient medium
parameters and cosmic ray distribution functions, and have reproduced
the plots of growth rate {\textit{vs}} wavenumber in Bell (2004) and
Blasi \& Amato (2008). An example is shown in Figure
\ref{growthrates.eps}, which plots growth rate vs wavenumber at fixed
cosmic ray flux, ion density, and magnetic fieldstrength for three
different temperatures: $T=10^4$\,K, $T=10^6$\,K, $T=10^7$\,K. The other
parameters, $B=3\,\mu G$, $n_i=1$\,cm$^{-3}$, $\ncr
v_D=10^4$\,cm$^{-2}$\,s$^{-1}$, are similar to those chosen by these
authors as representative of cosmic ray acceleration in a young
supernova remnant.

 The lowest temperature is essentially the cold plasma
result\footnote{At sufficiently low ISM temperatures ion-neutral
damping of the fluctuations must also be considered; (Zweibel \& Shull
1982, Reville et al. 2008b), but that is beyond the scope of this
paper.}.  The growth rate peaks at $\kb$ and then plunges, but remains
positive as $k$ increases. The $T=10^6$ curve is similar to the
$T=10^4$\,K curve, except that the peak growth rate is reduced
and occurs at a slightly smaller $k$, and the positive tail is
damped. The $T=10^7$\,K curve is markedly different. The
wavelength of the fastest growing mode is longer, and the peak growth
rate is lower, than at $T=10^6$\,K.  The instability cuts off abruptly
at a longer wavelength than in the $T=10^4$\,K and $T=10^6$\,K cases, then
re-emerges in a short interval of $k$ before disappearing again.  We
explain these features with an analytical treatment in the following
two subsections.

\begin{figure}[h!]
\begin{center}
\includegraphics[height=50mm]{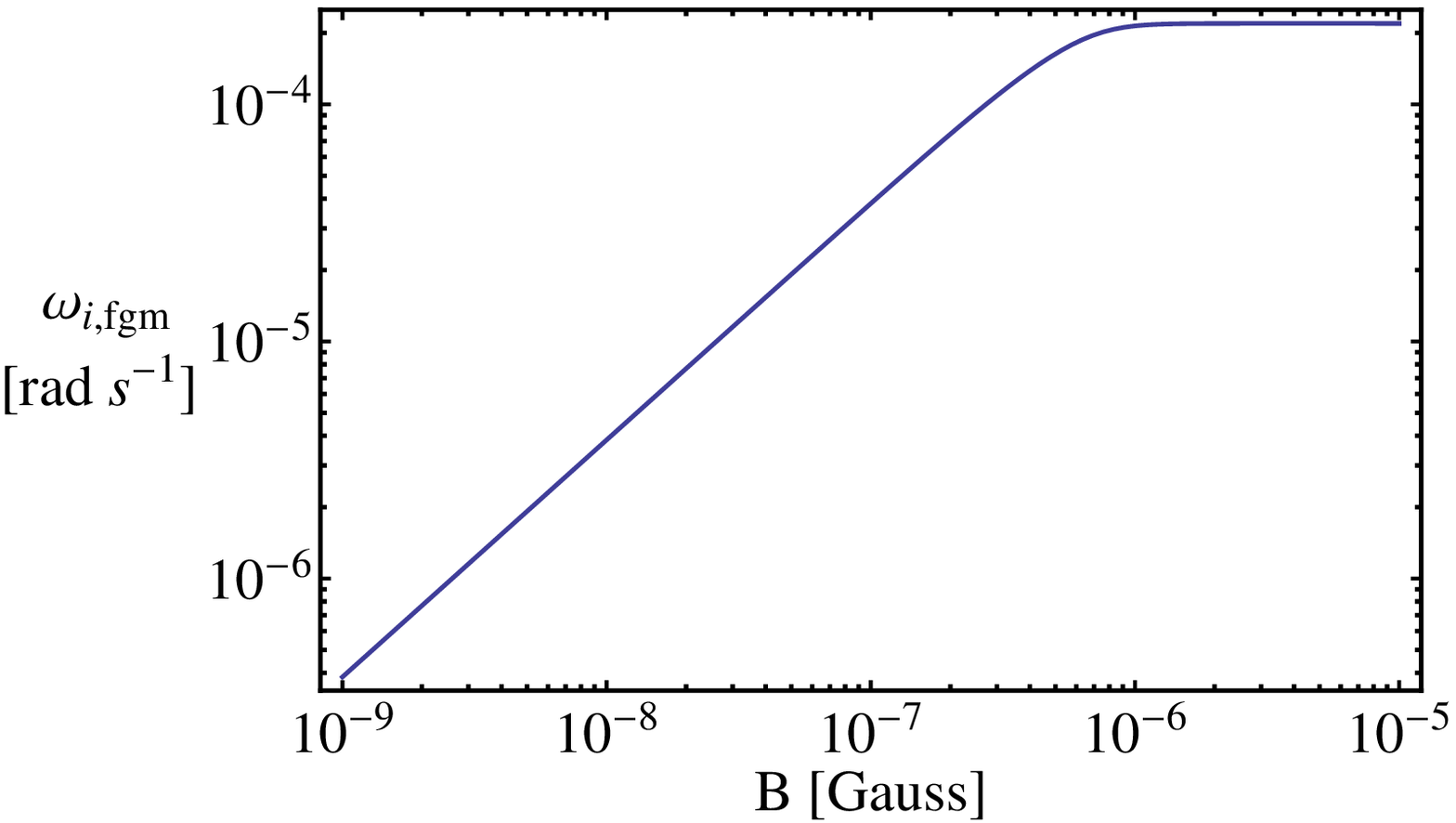}
\includegraphics[height=50mm]{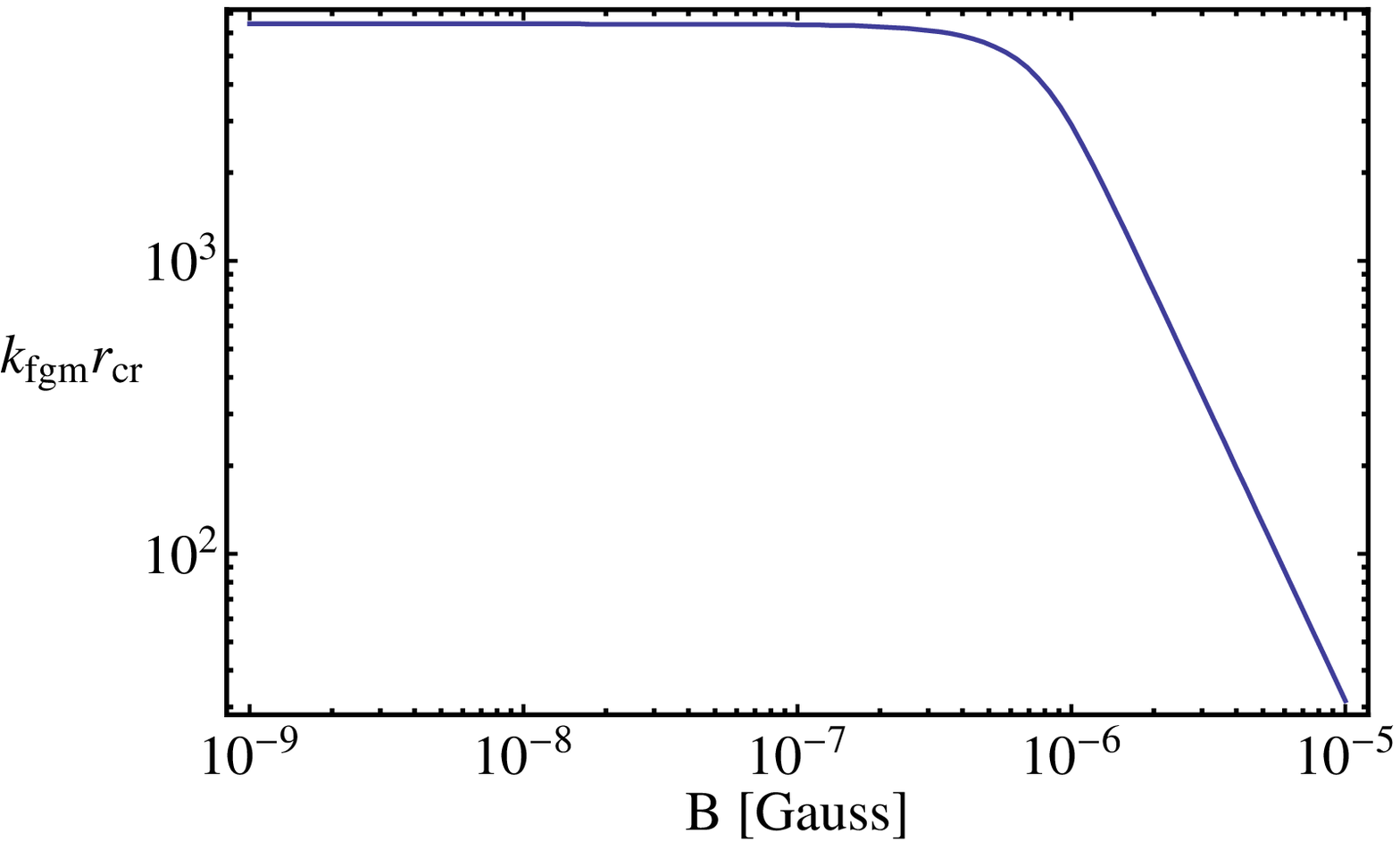}
\caption{\label{fgm}Left panel: maximum growthrate of the streaming
instability as a function of magnetic fieldstrength. Right panel:
wavenumber $k$ of the fastest growing mode. Here, $T=10^4$\,K and the
other parameters are as in Figure \ref{growthrates.eps}.}
\end{center}
\end{figure}

The effect of magnetic fieldstrength on the fastest growing mode is
shown in Figures (\ref{fgm}a,b).  Figure (\ref{fgm}a) plots the
maximum growthrate $\om_{i,fgm}$ {\textit{vs}} $B$ at $T=10^4$\,K,
with other parameters as in Figure (\ref{growthrates.eps}).  For
$B\le 1\,\mu G$, $\om_{i,fgm}$ increases linearly with $B$. At larger
$B$, $\om_{i,fgm}$ is independent of $B$. Different behavior is seen
for $k_{fgm}$: the wavenumber of the fastest growing mode,
plotted in Figure (\ref{fgm}b), is independent of $B$ at low
fieldstrength (here, for $B\le 1\,\mu G$) and decreases as
$B^{-1}$ at larger $B$. These features, too, are derived in the
next two subsections.
\subsection{Analytical Results}
\subsubsection{Standard case}\label{s:stdCase}

We recover the essential form of the Bell instability from
eqn. (\ref{fulldr}) by setting $\zeta_r=-1$ and taking the zero
temperature limit of the right hand side. Using $Z(z)\rightarrow -1/z
$; $\vert z\vert\gg 1$, eqn. (\ref{fulldr}) becomes
\begin{equation}\label{largez1}
\frac{c^2k^2}{\om^2} + \omci\frac{\ncr}{n_i}\frac{\left(\om-kv_D\right)}{\om^2}\frac{c^2}{v_A^2}=
-\frac{\ompi^2}{\om(\omci+\om)}-
\frac{\ompe^2}{\om(\omce+\om)}.
\end{equation}

We now assume $\om\ll\omci$, approximate $(\omci+\om)^{-1}$ by $\omci^{-1}(1-\om/\omci)$, and use $\ompe^2/\omce=-\ompi^2/\omci$.
Multiplying the resulting dispersion relation by $\om^2v_A^2/c^2$ yields 
\begin{equation}\label{largez}
k^2v_A^2 +\omci\frac{\ncr}{n_i}\left(\om-kv_D\right)=\om^2.
\end{equation}
The solution to eqn. (\ref{largez}) is 
\begin{equation}\label{largezroots}
\om=\frac{\omci}{2}\frac{\ncr}{n_i}\pm\left[\left(\frac{\omci}{2}\frac{\ncr}{n_i}\right)^2-\omci\frac{\ncr}{n_i}kv_D+k^2v_A^2\right]^{1/2}.
\end{equation}
The fastest growing mode occurs at the Bell wavenumber $\kb$
\begin{equation}\label{kbell}
\kb\equiv\frac{\omci}{2}\frac{\ncr}{n_i}\frac{v_D}{v_A^2}
\end{equation}
and has
\begin{eqnarray}\label{omegabell}
\om(\kb)\equiv\om_{Bell}& = & \frac{\omci}{2}\frac{\ncr}{n_i}\left[1+\frac{v_D}{v_A}\left(\frac{v_A^2}{v_D^2}-1\right)^{1/2}\right]
\nonumber \\
& \rightarrow & i \frac{\omci}{2}\frac{\ncr}{n_i}\frac{v_D}{v_A},
\end{eqnarray}
where the limiting expression on the last line holds for
$v_D/v_A\gg 1$.

Equation (\ref{omegabell}) shows that instability requires
$v_D/v_A>1$. This is also the threshold for the classical resonant
streaming instability. When the cosmic ray flux is too low to satisfy
eqn. (\ref{bellineq}), the dominant instabilities are resonant
instabilities of Alfv\'en waves. The peak growth rate occurs at
wavenumbers which resonate with cosmic rays near the mean energy,
$k\sim r_{cr}^{-1}$, and is of order $\omci(\ncr/n_i)(v_D/v_A-1)$
(Kulsrud \& Cesarsky 1971). For $v_D/v_A\gg 1$, this expression agrees
to within a factor of order unity with $\om_{Bell}$, but the two
cannot be used simultaneously because they apply for opposite cases of
the inequality (\ref{bellineq}).
  
At fluxes which satisfy eqn. (\ref{bellineq}), the assumptions made in
deriving the classical resonant growth rate - that the underlying
waves are Alfven waves and that the growth time is much longer than
the wave period - are incorrect (Zweibel 1979, 2003, Achterberg
1983). At wavenumbers $k < r_{cr}^{-1}$, the growth rate is of order
$\omci(\ncr/n_i)(v_D/v_A-1)^{1/2}$ for $v_D$ slightly greater than
$v_A$ and peaks at $\omci(\ncr v_D/n_ic)^{1/2}$ for $v_D\gg v_A$
(Zweibel 2003).  Comparing this expression to $\om_{Bell}$, we see
that $\om_{res}/\om_{Bell}\sim (n_i
v_A^2/\ncr c v_D)^{1/2}$.  From eqn. (\ref{bellineq}), we see that
whenever the Bell instability operates, its growth rate exceeds the
growth rate of the resonant instability, and the growth rate of the resonant instability
is lower than predicted by the classical theory.

At $k>2\kb$, the nonresonant instability is stabilized by
magnetic tension. Comparison
with Figure \ref{growthrates.eps} shows that although the growthrate
plunges to low levels at this value of $k$, the instability does not
completely disappear, as is predicted by eqn. (\ref{largez}).  The
$\om_i\propto k^{-1}$ tail seen in Figure \ref{growthrates.eps} can be
recovered by using the full $\zeta_r$; for $kr_{cr}\gg 1$,
$\zeta_r\sim -1 +i/kr_{cr}$.

The Bell instability is important in shock acceleration if the growth
time is shorter than the time it takes the shock to travel through the
layer within which the cosmic rays are confined. For acceleration at
the maximum rate, the cosmic ray diffuse according to the Bohm formula
with diffusion coefficient $D\sim cr_{cr}$, which sets the convection
time across the layer as $(\omci v_D^2/c^2)^{-1}$. The Bell
instability will be able to grow if $(\ncr v_D/n_iv_A)>v_D^2/c^2$.

\subsubsection{Warm ions, cold electrons}\label{s:wice}

We now imagine decreasing $B$ or increasing $T$ such that
$kv_i/\vert\omci+\om\vert$ is less than unity but not infinitesimal.
In this case, $Z(z_i)\sim -1/z- 1/2z^3+i\sqrt{\pi}e^{-z^2}$. The
imaginary part represents ions for which the Doppler shifted wave
frequency $\om - kv=-\omci$; i.e. ions in cyclotron resonance. The
$-1/2z^3$ term represents the finite gyroradius of the ions, which
partially decouples them from the field. Using this approximation,
eqn. (\ref{fulldr}) becomes
\begin{equation}\label{wicedr}
\om^2-\om\left(\frac{k^2v_i^2}{2\omci}+\omci\frac{\ncr}{n_i}-i\sqrt{\pi}\frac{\omci^2}{kv_i}e^{-\frac{\omci^2}{k^2v_i^2}}
\right)-k^2v_A^2+\omci\frac{\ncr}{n_i}kv_D=0.
\end{equation}
Comparison of eqns. (\ref{largez}) and (\ref{wicedr}) shows that the
thermal ions have two effects: cyclotron resonance, which is
represented by the imaginary term, and a pressure-like effect due to
the finite ion gyroradius, which increases the wave speed and is known
as ion gyroviscosity.  In the absence of cosmic rays, the dispersion
relation agrees with Foote \& Kulsrud (1979), and in the limit
$kr_{cr}\gg 1$, with Reville et al. (2008b),\footnote{The sign of the
second term on the left hand side of eqn. (3) in Reville et
al. (2008b) is incorrect.} except that ion cyclotron damping is
neglected in both papers.

\begin{figure}[h!]
\begin{center}
\includegraphics[height=45mm]{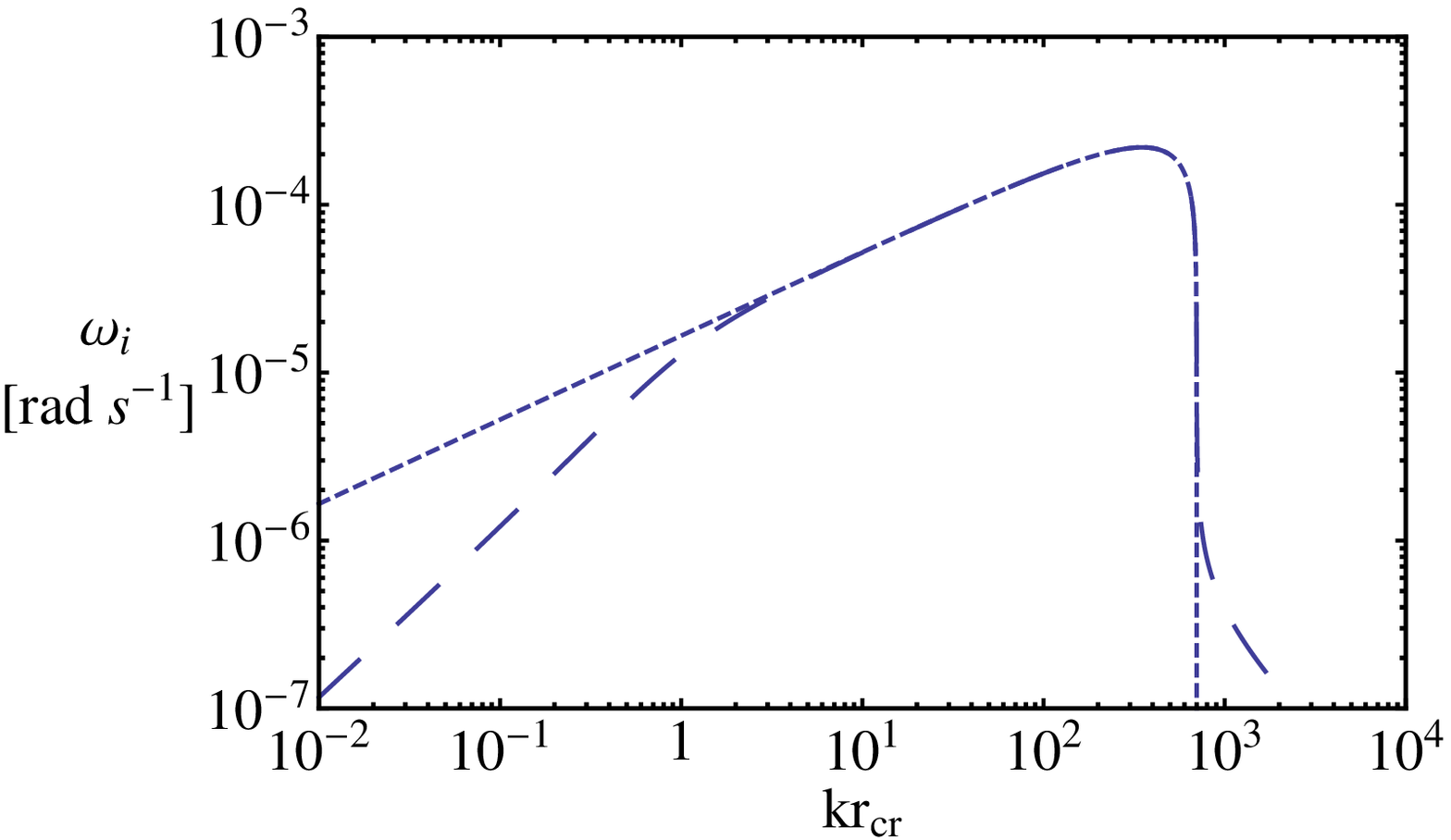}
\includegraphics[height=45mm]{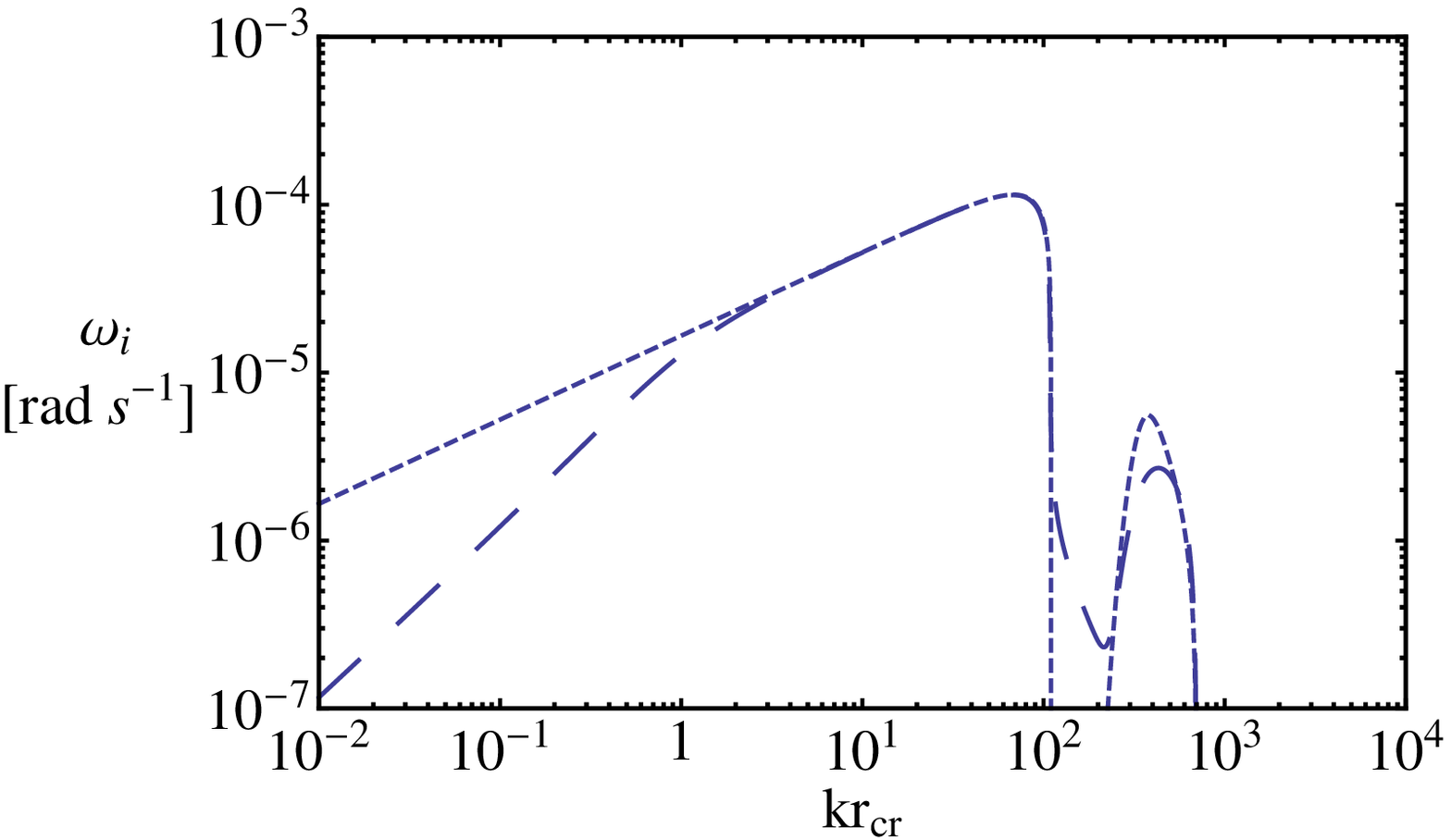}
\caption{\label{comparison.eps}Left panel: Growth rate $\om_i$ vs
scaled wavenumber $kr_{cr}$ as computed from the full dispersion
relation eqn. (\ref{fulldr}) - long dashed line -, and the
approximation (\ref{wicedr}) - short dashed line for $T=10^3$\,K and the
other parameters as in eqn. (\ref{growthrates.eps}). Right panel: same
for $T=10^7$\,K.}
\end{center}
\end{figure}

Equation (\ref{wicedr}) represents the behavior of the growth rate
quite well. This is shown in Figure (\ref{comparison.eps}), which
compares $\om_i$ calculated from eqns. (\ref{fulldr}) and
(\ref{wicedr}).

In the left panel, $T=10^3$\,K, thermal effects are unimportant and the
differences between the two curves are entirely due to the value
adopted for $\zeta_r$; taking $\zeta_r=-1$ omits the effect of
resonant particles at high $k$ and underestimates the cosmic ray
response at low $k$, thereby overestimating $\om_i$. In the right
panel, $T=10^7$\,K, thermal effects are important enough to modify the
instability (see Figure \ref{growthrates.eps} and below) but the
analytical dispersion relation successfully reproduces the main
features. Again, setting $\zeta_r=-1$ overestimates the growthrate at
low $k$.

It can be shown from eqn. (\ref{wicedr}) that when
\begin{equation}\label{wice}
\frac{v_A}{v_i} < \left(\frac{\ncr v_D}{n_iv_i}\right)^{1/3},
\end{equation}
the fastest growing mode is determined by competition between the
drift term $\omci kv_D\ncr/n_i$ and the finite gyroradius term
rather than the drift term and the magnetic tension term. This
condition is equivalent to eqn. (\ref{revilleineq}).  The wavenumber
$k_{wice}$ of the fastest growing mode in this regime is
\begin{equation}\label{kmax}
k_{wice}\sim\frac{\omci}{v_i}\left(\frac{\ncr}{n_i}\frac{v_D}{v_i}\right)^{1/3}.
\end{equation}

Equation (\ref{kmax}) neglects cyclotron damping, and thus is valid
only for $\kw r_i\sim (\ncr v_D/n_iv_i)^{1/3} < 1$. As
$k_{wice}r_i\rightarrow 1$ cyclotron damping becomes large, and the
instability shuts off. A further constraint is that the instability be
nonresonant: $\kw r_{cr}\sim (c/v_i) \kw r_i > 1$. Both requirements
can be written as temperature and density dependent limits on $\ncr
v_D$; we return to them in \S\ref{ss:regimes} (eqns. \ref{bm},
\ref{bt}).

The 
growth rate $\om_{wice}$ corresponding to $k_{wice}$
is
\begin{equation}\label{omax}
\om_{wice}\sim \omci\left(\frac{\ncr}{n_i}\frac{v_D}{v_i}\right)^{2/3},
\end{equation}
in agreement with Reville et al. (2008b).

It can be shown from eqns. (\ref{omegabell}), (\ref{wice}), and
(\ref{omax}) that $\om_{wice}/\om_{Bell} < 1$.  At the limits of
validity of the warm ion approximation, which is
$k_{wice}v_i/\omci=1$, $\om_{wice}/\om_{Bell}=v_A/v_i$.  The
suppression of the growth rate is due not to thermal ion cyclotron
damping, which is weak for $\omci/kv_i\gg 1$, but to the restoring
force exerted by the warm ions.  Cyclotron damping does, however, come
into play at shorter wavelength, obliterating the resonant tail of the
instability. This happens roughly where
$(kv_i/\omci)e^{-(\omci/kv_i)^2} > \ncr v_D/n_i c$. For the cosmic ray
flux and ion density assumed in Figures \ref{growthrates.eps} -
\ref{comparison.eps}, this occurs at $kv_i/\omci\sim 0.27$, or
$kc/\omci\sim 9\times 10^5 T^{-1/2}$. This is consistent with the
behavior shown in Figure \ref{growthrates.eps}.

At $T=10^7$\,K, something more complicated is going on: the instability
growth rate decreases sharply after peaking near $kc/\omci\sim 10^2$,
as predicted by eqn. (\ref{kmax}), but then has a brief
resurgence. This is because the resonant ion cyclotron term overwhelms
the stabilizing gyroviscous term for $\omci/kv_i\sim 3/2$, removing, in
a small band of $k$ space, gyroviscous stabilization.

As we discussed in \S 2.2.1, the instability can only efficiently
amplify magnetic fields at a shock if its growth time is faster than
the convection time $(\omci v_D^2/c^2)^{-1}$. From eqn. (\ref{omax})
we see that this requires
\begin{equation}\label{fast}
\omci\left(\frac{\ncr}{n_i}\frac{v_D}{v_i}\right)^{2/3} > \omci\frac{v_D^2}{c^2}.
\end{equation}
The criterion (\ref{fast}) is independent of magnetic fieldstrength,
and depends only on the cosmic ray flux, drift speed, and the density
and temperature of the ambient medium.

\subsubsection{Hot ions}

When $kr_i\ge 1$, eqn. (\ref{wicedr}) becomes invalid. In this
limit, the argument of the plasma dispersion function in
eqn. (\ref{fulldr}) becomes large, and $Z(z)\approx -z^{-1} +
i\sqrt{\pi}$. Physically, this means the ions are responding very
little to the perturbation. Ion cyclotron damping is also weak,
because the slope of the distribution function is small at the
resonant velocities.

Under these conditions, the instability, if it exists at all, is due
entirely to properties of the electron distribution function. In
deriving the cosmic ray response function $\zeta_r$ used in
eqn. (\ref{fulldr}) and plotted in Figure (\ref{zeta.eps}), we assumed
the electrons are cold and a fraction $\ncr/n_i$ of them are drifting
with the cosmic rays. As long as $v_D/v_e$ is sufficiently large, the
electrons are unstable not only to the electromagnetic streaming
instability considered here, but also to the much faster growing
electrostatic instabilities discussed in \S\ref{s:setup}.  On the
other hand, if the electrons were all drifting at speed $\ncr v_D/v_i$
both the electromagnetic and electrostatic instabilities would be
stabilized.

Assuming the two peaked electron distribution, it can be shown that
the instability growth rate is bounded above by $\omci (\ncr
v_D/n_iv_i)$. This is generally less than $\om_{wice}$ defined in
eqn. (\ref{omax}), showing that these very short wavelength
instabilities are not as important as the thermally modified or
standard Bell instabilities. At even shorter wavelengths, such that
$kr_e > 1$, the derivation of $\zeta_r$ becomes invalid.  In view of
our uncertainty about the electron distribution function, we have
pursued the hot ion case no further.

\section{Applications}\label{s:applications}

\subsection{Instability regimes}\label{ss:regimes}
We begin by summarizing the different regimes of the streaming
instability as functions of magnetic fieldstrength $B$ and cosmic ray
flux $\ncr v_D$. These regimes were introduced without proof in
\S\ref{s:introduction} and depicted schematically in Figure
(\ref{planesketch.eps}).  The criteria used to delineate these regimes
are approximate, but as Figures \ref{fgm}a,b indicate, the transitions
between regimes are fairly sharp.

According to eqn. (\ref{bellineq}), the condition that the Bell
instability be nonresonant, i.e. that $\kb\rcr > 1$, is
\begin{equation}\label{bs}
B < B_S\equiv 8.7\times 10^{-7}(\ncr v_D)^{1/2},
\end{equation}
where here and below $\ncr v_D$ is given in units of cm$^{-2}$
s$^{-1}$ and $B$ is in G.  The condition that thermal effects modify
the Bell instability such that the wavelength of the fastest growing
mode is at $k\sim k_{wice}$ (eqn. \ref{kbell}) rather than $k\sim\kb$
(eqn. \ref{kmax}) is
\begin{equation}\label{bm}
B < B_M\equiv 2.3\times 10^{-9}T^{1/3}n_i^{1/6}(\ncr v_D)^{1/3}.
\end{equation}
The condition for the thermally modified Bell instability to be
nonresonant is $k_{wice}\rcr > 1$. At the same time, the thermal ions
must be magnetized at $k=k_{wice}$; $k_{wice}r_i < 1$. These
conditions limit $\ncr v_D$ to the range
\begin{equation}\label{symbolflux}
n_i\frac{v_i^4}{c^3} < \ncr v_D < n_i v_i,
\end{equation}
 or numerically
\begin{equation}\label{flux}
3\times 10^{-16}n_iT^2 < \ncr v_D < 10^4 n_iT^{1/2}. 
\end{equation}
Finally, the condition that the ions be magnetized at $k=\kb$, $\kb
r_i < 1$, is $\ncr v_D < n_i v_A^2/v_i$, or
\begin{equation}\label{bt}
B > B_T\equiv 5\times 10^{-10}T^{1/4}(\ncr v_D)^{1/2}.
\end{equation}

\begin{figure}[h!]
\begin{center}
\includegraphics[height=50mm]{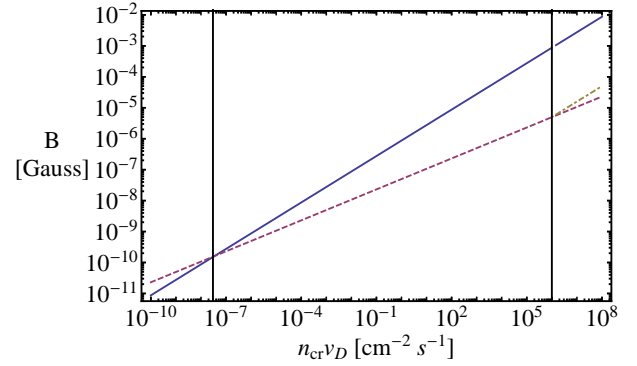}
\caption{\label{ismregimes.eps} Log-log plot of the ($\ncr v_D, B$)
plane with various regimes delineated. The quantities $B_S$, $B_M$,
and $B_T$ defined in eqns. (\ref{bs}, (\ref{bm}), and (\ref{bt})
appear as lines with slope $1/2$, $1/3$, and $1/2$, respectively. The
flux limits defined in eqn. (\ref{symbolflux}) appear as vertical
lines. The plasma density is $n_i=1$\,cm$^{-3}$ and the temperature is
$T=10^4$\,K.}
\end{center}
\end{figure}

\begin{figure}[h!]
\begin{center}
\includegraphics[height=50mm]{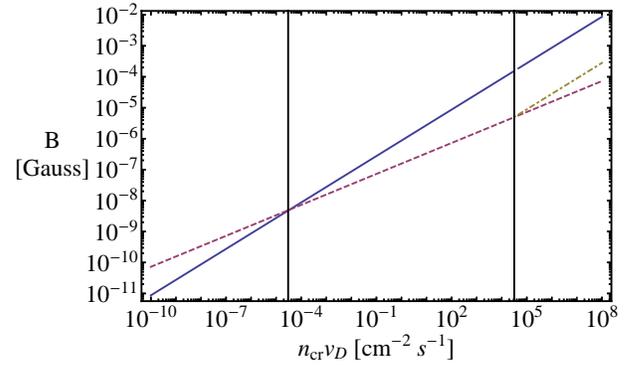}
\caption{\label{icmregimes.eps} Same as Figure \ref{ismregimes.eps}
except $n_i=10^{-3}$\,cm$^{-3}$, $T=10^7$\,K}.
\end{center}
\end{figure}

Equations (\ref{bs}) - (\ref{bt}) are plotted on the ($\ncr v_D$, $B$)
plane) in Figures \ref{ismregimes.eps} and \ref{icmregimes.eps}.
Because eqns. (\ref{bm}) - (\ref{bt}) depend on $n_i$ and $T$, we give
two versions of the plot. Figure \ref{ismregimes.eps} represents the
interstellar medium; $n_i=1$\,cm$^{-3}$, $T=10^4$\,K. Figure
\ref{icmregimes.eps} represents the intracluster medium;
$n_i=10^{-3}$\,cm$^{-3}$, $T=10^7$\,K.

For $\ncr v_D < n_iv_i^4/c^3$, $B_S<B_M$. This is also the condition
$k_{wice}\rcr < 1$ (eqn. \ref{flux}), and is represented by the
leftmost vertical line on the plot. To the left of this line, there
can be no nonresonant instability; since $B_M > B_S$ the instability
would take the thermally modified form, but for $B>B_M$ the field is
too large and for $B<B_M$ the flux is too low. Streaming instability
exists, but it is resonant. In the cold plasma limit, the Bell
instability exists for any $B<B_S$, no matter how small.

For $\ncr v_D > n_i v_i^4/c^3$, $B_M < B_S$. For $B>B_S$, the field is
too large for nonresonant instability (this is the case in the
interstellar medium, away from cosmic ray sources). For $B < B_S$, the
Bell instability operates in standard form as long as $B$ exceeds
$B_M$ and $B_T$. Although $B_M<B_T$ is theoretically possible, it
requires $\ncr v_D > n_i v_i$, which is rather extreme. If we confine
ourselves to $\ncr v_D < n_i v_i$ (represented by the rightmost
vertical line, defined by $\kw\ri = 1$then the Bell instability
operates for $B_M < B < B_S$. For $B < B_M$, there is nonresonant
instability as long as $\ncr v_D$ is to the left of the vertical
line.
To the
right of this line, the ions are unmagnetized.  As we have argued, in
this case the instability is controlled primarily by the electrons,
and for $kr_i\sim 1$, ion cyclotron damping is strong.

\begin{figure}[h!]
\begin{center}
\includegraphics[height=80mm]{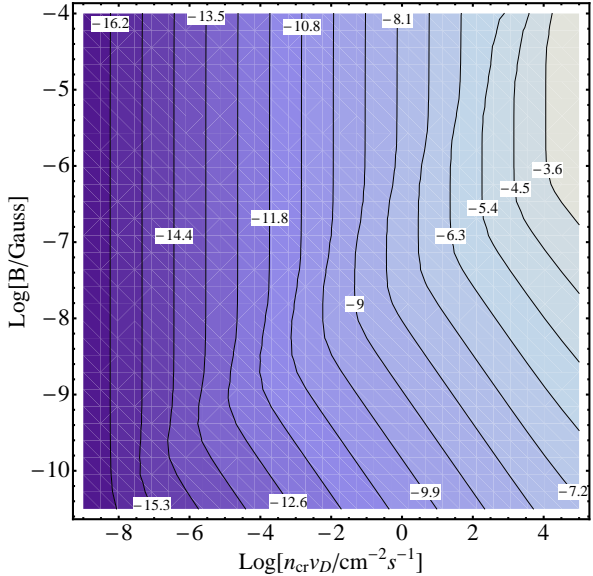}
\caption{Contour plot of the maximum growth rate
($\omega_{fgm}$) of Eqn~(\ref{fulldr}) plotted as a function of the
logarithm of the cosmic-ray flux $n_{cr} v_D$ and the logarithm of the
magnetic-field strength, $B$.  The growth rate does not change with
$B$ in the region where the resonant and Bell Instability
dominate, but at smaller magnetic-field strengths, $\omega_{fgm}$
starts to decrease linearly with $B$.  At relatively large magnetic
field strengths and low cosmic-ray fluxes ($B \sim 10^{-6}$\,G and
$n_{cr}v_D \sim 2.5$\,cm$^{-2}$\,s$^{-1}$), the growth rate falls
slightly in the transition between the non-resonant Bell Instability
and the resonant streaming instability, giving a slight curvature to
the contours there.
\label{maxGrowthRateContourPlot}}
\end{center}
\end{figure}

In summary, nonresonant instabilities exist in the range of cosmic ray
fluxes given by eqn. (\ref{flux}). When $B$ is between $B_M$, defined
in eqn. (\ref{bm}), and $B_S$, defined in eqn. (\ref{bs}), the maximum
growthrate is independent of $B$. When $B < B_M$ the instability is
thermally modified, occurs at longer wavelength, and grows at a rate
proportional to $B$.  This is also shown in
Figure~\ref{maxGrowthRateContourPlot}, which is a contour plot of
the maximum growth rate in Equation~(\ref{fulldr}) as a function of
cosmic-ray flux and magnetic field. Towards the lower-right of the
plot, one can see the maximum growth rate decreasing, downward, with
decreasing $B$, but relatively constant above an approximately
diagonal line in ($n_{cr} v_D$,$B$) space from
$(10^{-10}$\,cm$^{-2}$\,s$^{-1}, 10^{-8}\,G)$ to
$(10^{-4.4}$\,cm$^{-2}$\,s$^{-1}, 10^{-5.8}\,G)$.

\subsection{Astrophysical settings}\label{ss:settings}

We now consider a few examples of specific astrophysical settings. In
order to smoothly represent the transition from the standard Bell
regime $B_M<B<B_S$ to the thermally modified regime $B<B_M$, we
replace $\om_{wice}$ by $\om_{Bell}B/B_M$, which agrees with
eqn. (\ref{omegabell}) up to a factor of order unity.
 
\subsubsection{Supernova remnants and superbubbles}\label{sss:snr}

First, we consider cosmic ray acceleration by a supernova driven shock
traveling at 10$^4$ km s$^{-1}$ through the interstellar medium. As in
Figure \ref{ismregimes.eps}a we take $n_i=1$\,cm$^{-3}$,
$T=10^4$\,K. According to eqn. (\ref{flux}), nonresonant
instability exists for $3\times 10^{-8}$\,cm$^{-2}$s$^{-1}
< \ncr v_D < 10^6$\,cm$^{-2}$s$^{-1}$. Taking $\ncr
v_D=10^4$\,cm$^{-2}$s$^{-1}$, as assumed in Figure
\ref{growthrates.eps}, we find from eqns. (\ref{bs}) and (\ref{bm})
that the standard Bell instability operates for $B$ between $1.1$ and
$87$\,$\mu G$, which encompasses most of the fieldstrengths measured in the
diffuse interstellar medium. The growth rate in the Bell regime is
$2.5\times 10^{-4}$ s$^{-1}$ and $2.5\times 10^{-4}(B/1.1\,\mu G)$ in
the thermally modified regime. The condition for efficient field
amplification at shocks, $\om > \omci (v_D/c)^2$, is $\om > 1.1\times
10^1 B$. Growth is amply fast in the Bell regime and also in the
thermally modified Bell regime, since the growth rate and advection
time scale in the same way with $B$. Of course, if $B$ is too small,
the cosmic ray acceleration time becomes long compared to the shock
evolution time. But even if $B\sim 10^{-9}G$, about an order of
magnitude less than the disordered field estimated by Rees (1987) to
arise from a superposition of plerion supernova remnants in the early
galaxy, the characteristic growth time of the instability is less than
a year, much faster than the timescale on which the remnant evolves.

On the other hand, if the shock propagates in a hot, low density medium such
as the superbubbles modelled by MacLow \& McCray (1988),
with $n_i = 3\times 10^{-3}$\,cm$^{-3}$, $T=3\times 10^6$\,K, then
according to eqn. (\ref{flux}), the nonresonant instability
could exist for $8.1\times 10^{-6}$\,cm$^{-2}$s$^{-1} < \ncr v_D <
5.2\times 10^4$\,cm$^{-2}$s$^{-1}$. If the cosmic ray injection
efficiency were the same as at higher densities, $\ncr/n_i = 10^{-5}$,
eqns. (\ref{bs}) and (\ref{bm}) show that the standard Bell
instability operates between 0.39 and 4.7\,$\mu$G. The upper limit is
only slightly less than the rms Galactic field. The growth rate in the
Bell regime, $1.4\times 10^{-5}$, exceeds the
advection rate through the cosmic ray scattering layer only for
$B < 1.3\mu G$, suggesting
that nonresonant instability is less important for shock acceleration
in a low density medium than in a high density medium. The injection
efficiency, rather than thermal effects, is the deciding factor: increasing
$\ncr/n_i$ above 10$^{-5}$ would enhance the growth rate in proportion.
These results imply that nonresonant instabilities might not occur for
cosmic ray acceleration in superbubbles, where the combined effects of
many hot star winds and explosions make the density low. 

Denser superbubbles have
been observed (Dunne et al. 2001). Increasing $n_i$ to $3\times 10^{-2}$ cm$^{-3}$ while leaving the other parameters
the same would increase $\om_{Bell}$ to $3.8\times 10^{-5}$ s$^{-1}$ and change
the range in which the standard Bell Instability operates to $1.2\mu G < B < 15\mu G$.
The RMS Galactic field is estimated to be
$\sim 5.5\mu G$ (Ferri\`ere 2001), so the conditions for instability should be satisfied. The instability growth rate exceeds 
the advection rate through the acceleration
layer as long as $B< 4.1G$.

\subsubsection{Shocks in galaxy clusters}\label{clusters}
Next, we consider acceleration at shocks in the intracluster
medium. We take $n_i\sim 10^{-3}$\,cm$^{-3}$, $T=10^7$\,K.  According to
eqn. (\ref{flux}), nonresonant instability exists for $3\times 10^{-5}
$\,cm$^{-2}$s$^{-1} < \ncr v_D < 3.1\times 10^4$\,cm$^{-2}$\,s$^{-1}$.  If
acceleration occurs at the same efficiency assumed for galactic
supernova remnants then $\ncr = 10^{-8}$\,cm$^{-3}$; assuming $v_D\sim
v_S\sim 1000$ km s$^{-1}$, then $B_S$, the maximum fieldstrength for
nonresonant instability, is $8.7\times 10^{-7}G$. This is slightly
less than the fields inferred in galaxy clusters (e.g., Govoni \&
Feretti 2004), but the parameters are too uncertain to rule out
resonant instability. According to eqn. (\ref{omegabell}), the growth
rate in the Bell regime is $7.5\times 10^{-7}$ s$^{-1}$. With these
same parameters, $B_M=1.6\times 10^{-7}G$; below this value the
instability is thermally modified and grows at the rate
$7.5\times
10^{-7}(B/.16\,\mu G)$ s $^{-1}$.  Due to the high temperature, the
range in which the standard Bell instability operates without thermal
effects is quite small, but the growth rate of the thermally modified
instability is still fast. The condition for efficient field
amplification at shocks, $\om > \omci (v_D/c)^2$, is $\om > 1.1\times
10^{-1} B$. Although the instability growth rates are lower than for
supernova remnants, they are still large enough to satisfy this
condition in both the standard and thermally modified nonresonant
regimes. Thus, it appears that nonresonant instabilities could play a
role in shock acceleration, and could amplify magnetic fields, in
galaxy clusters. This could make intergalactic shocks a favorable
environment for acceleration of ultra high energy cosmic rays.

\subsubsection{Unconfined galactic cosmic rays}

As our final example, we consider leakage of cosmic rays from galaxies
into the intergalactic medium, which we take to have density and
temperature $n_i=10^{-6}$\,cm$^{-3}$, $T=10^6$\,K (Richter et al. 2008). 
From eqn. (\ref{flux}),
nonresonant instabilities can be excited by cosmic ray fluxes between
$3\times 10^{-10}$ and $10$\,cm$^{-2}$\,s$^{-1}$.  In the local
interstellar medium, $\ncr\sim 10^{-9}$\,cm$^{-3}$, while $v_D$ is
roughly the scale height of cosmic rays divided by their confinement
time in the galaxy, or about 100 km s$^{-1}$. This gives a galactic
flux $\ncr v_D\sim 10^{-2}$\,cm$^{-2}$\,s$^{-1}$. Assuming cosmic rays
emanate isotropically from a characteristic galaxy size $R_g$, we
write $\ncr v_D\sim 10^{-2}(R_g/R)^2 (L_{cr}/L_{crMW})$. From
eqns. (\ref{bs}) and (\ref{bm}), we find $B_S= 8.7\times
10^{-8}(R_g/R) (L_{cr}/L_{crMW})^{1/2}$ while $B_M = 7.3\times 10^{-9}
(R_g/R)^{2/3}(L_{cr}/L_{crMW})^{1/3}$. These values suggest that
nonresonant instabilities could be excited in the intergalactic medium
even if the fields are weaker than the $10^{-9}-10^{-10}$\,G range often
cited as upper limits (Kulsrud \& Zweibel 2008). The growth rates,
however, are rather slow. For example, if
$R/R_g = 10$, $\om_{Bell}=2.5\times
10^{-9}(L_{cr}/L_{crMW})$. In this case, $B_M = 1.6\times 10^{-9}$ G;
for $B=10^{-10}$G, the maximum growth rate is about $4.1\times
10^{-10} (L_{cr}/L_{crMW})$ s$^{-1}$. Still, although the growth time
exceeds 10$^3$ years, this is much shorter than any reasonable cosmic
ray convection time. Therefore, nonresonant instabilities could be
excited by cosmic rays from ordinary galaxies in the intergalactic
medium at large. They could amplify intergalactic magnetic fields, and
could heat the plasma.

\section{Summary}\label{s:summary} 

Nonresonant instability driven by cosmic ray streaming has emerged as
a strong candidate for amplification of magnetic fields in
environments such as strong shock waves, where the cosmic ray flux is
large (Bell 2004). When the flux is high enough and/or the magnetic
field is low enough, that eqn. (\ref{bellineq}) is violated, the
nonresonant instability replaces the classical resonant streaming
instability as the dominant electromagnetic instability generated by
cosmic rays. Although the instability scale length predicted by linear
theory is small even compared to the cosmic ray gyroradius $r_{cr}$,
nonlinear simulations suggest that as the amplitude of the instability
grows it generates fluctations at larger scales. This can increase the
energy to which particles are accelerated in shocks.

Cosmic ray acceleration and magnetic field growth are both of interest
in a variety of environments, including young galaxies which may be
actively forming stars but have not yet built up magnetic fields,
shocks in galaxy clusters, and the intergalactic medium at large. In
this paper we have carried out a parameter study of nonresonant
instabilities including ion thermal effects. We solved the full
dispersion relation (\ref{fulldr}) numerically and verified that a
simple analytical approximation, eqn. (\ref{wicedr}), is quite
accurate in the wavenumber regime of interest. We corroborated the
criterion of Reville et al. (2008b) for when ion gyroviscosity reduces
the instability growth rate and shifts it to longer wavelength. We
showed that ion cyclotron damping cuts off the instability at short
wavelengths and argued that at wavelengths short enough that the ions
are unmagnetized the instability depends only on the electron
distribution function, the prediction of which is beyond the scope of
this paper.

The joint requirements that the instability wavelength be much less
than the cosmic ray gyroradius but much more than the thermal
ion gyroradius limits the range of fluxes which excite nonresonant
instability to $n_i v_i^4/c^3 < \ncr v_D < n_i v_i$. In practical
terms, this range is large and accommodates most cases of interest.
Within the unstable range, there is a ``strong field" regime in which
all streaming instabilities are resonant, an ``intermediate" regime in
which nonresonant instability in the form derived by Bell dominates,
and a ``weak field" regime in which the instability is thermally
modified.  In the Bell regime the maximum growth rate is independent of $B$
but in the thermally modified regime it depends linearly on $B$. Young
galactic supernova remnants are generally in the intermediate regime
unless the ambient medium is hot and rarefied (like the interior of a
superbubble), in which case the instability is weakened.  Generally,
if $B$ is in the nanogauss range, the growthrates are fast enough for
nonresonant instability to be a potential source of magnetic field
amplification in weakly magnetized interstellar and intergalactic
gas. At much lower fieldstrengths, the instability is too slow to be
of interest, but other instabilities, such as Weibel modes, could be
an important ingredient in magnetogenesis (Medvedev et al.
2006).

Although nonresonant instabilities amplify magnetic fields on rather
small scales - much smaller than the eddy scales characteristic of
interstellar and intergalactic turbulence - they should not be ignored
in discussions of magnetogenesis.  Because only the right circularly
polarized modes are unstable, nonresonant instabilities are a source
of magnetic helicity on scales at which the background magnetic field
is coherent. Magnetic helicity is thought to be a key ingredient in
the growth of large scale magnetic fields from small scale
fluctuations (Pouquet et al. 1976).  Cosmic ray generated fluctuations
could be important in driving an inverse cascade of magnetic power to
longer wavelengths, and could prevent the pileup of power at short
wavelengths that currently confounds interstellar and intergalactic dynamo theories.

\acknowledgements We are happy to acknowledge useful discussions with P. Blasi, J. Kirk, and B. Reville,
and comments by the referee. Support was provided by NSF
grants AST 0507367, PHY 0821899, and AST 0907837 to the University of Wisconsin.

\appendix
\section{Appendix: Electron Distribution Function}

Here we briefly consider the constraints on the electron distribution function $f_e$

One way or another, the cosmic ray current must be cancelled: an
uncompensated cosmic ray current $e\ncr v_D$ flowing in a channel of
width $L_{pc}$ measured in parsecs generates a magnetic field $B\sim
0.5\ncr v_DL_{pc}$\,G. Even the galactic flux of
$10^{-2}$\,cm$^{-2}$\,s$^{-1}$ with $L_{pc}=1$ would generate a 5\,mG
field. Since cosmic rays are ion dominated, thermal electrons must
cancel their flux.

When an electron beam drifts with respect to the bulk plasma, it can
excite rapidly growing electrostatic instabilities which tend to
redistribute electron momentum and bring the system to a state of
marginal stability. Langmuir waves (also called plasma oscillations)
with wavenumber $\ompe/v$ ($\ompe$ is the electron plasma frequency
$(4\pi n_ee^2/m_e)^{1/2}$) are destabilized if $\partial f_e/\partial
v > 0$.  If the beam and bulk electrons have the same temperature
$T_e$ and the beam velocity $v_b$ much exceeds the electron thermal
velocity $v_e\equiv\sqrt{2k_BT_e/m_e}$ (which is necessary for
instability if the beam density $n_b$ is much less than the bulk
density $n_e$, the case of interest here), then the requirement for
stability is approximately
\begin{equation}\label{langmuir}
\frac{v_b}{v_e}
e^{-v_b^2/v_e^2} > \frac{n_b }{n_e }
\end{equation}
(e.g. Krall \& Trivelpiece 1973).  In shock acceleration, it is
sometimes assumed $n_b/n_e \sim 10^{-5}$, According to
eqn. (\ref{langmuir}), stability then requires $v_b/v_e <
3.5$. Assuming $T_e=T_i$ in the upstream plasma, beams associated with
shocks of Mach number $M<3.5\sqrt{m_i/m_e}$ are stable while shocks at
higher $M$ are unstable. In a 10$^4$\,K gas, the stability boundary is
at about 1500\,km\,s$^{-1}$. Thus, while the Langmuir instability is a
constraint for very fast shocks, it is probably irrelevant for older
supernova remnants, and in galaxy cluster accretion shocks or galactic
wind termination shocks, where the background gas is hot and the Mach
numbers are expected to be moderate. The fluxes associated with cosmic
ray escape from galaxies are probably also electrostatically stable.

Therefore, it appears that $f_e$ is not determined by stability
considerations alone, but depends on other factors such as the history
of the system and the source of cosmic rays. 

\end{document}